\documentclass[reprint, amsmath, amssymb, pra,]{revtex4-2}

\usepackage{amssymb}
\usepackage{caption}
\usepackage{graphicx}
\usepackage{enumitem}
\usepackage{booktabs}
\usepackage{dcolumn}
\usepackage{bm}
\usepackage{amsmath}
\usepackage{qcircuit}

\usepackage{todonotes} 

\usetikzlibrary{arrows}

\newcommand{\QMA}{{\sf QMA}}
\newcommand{\BQP}{{\sf BQP}}
\newcommand{\SAT}{{\sc 3-SAT}}
\newcommand{\NP}{{\sf{NP}}}

\def\LH{{\sc local Hamiltonian}}
\def\2LH{{\sc $2$-local Hamiltonian}}
\def\5LH{{\sc $5$-local Hamiltonian}}

\usepackage{bbold}

\newcommand{\bra}[1]{\langle#1|}
\newcommand{\ket}[1]{|#1\rangle}

\newcommand{\braket}[2]{\langle#1|#2\rangle}
\newcommand{\bydef}{\stackrel{\mathrm{def}}{=}}

\newtheorem{theorem}{Theorem}
\newtheorem{proof}{Proof}
\newtheorem{lemma}{Lemma}
\newtheorem{definition}{Definition}
\newtheorem{example}{Example}
\newtheorem{remark}{Remark}

\begin{document}

\title{Universal Variational Quantum Computation}
\author{Jacob Biamonte}%
 \email{j.biamonte@skoltech.ru}
  \homepage{http://quantum.skoltech.ru}
\affiliation{Skolkovo Institute of Science and Technology, 3 Nobel Street, Moscow, Russia 121205}


\date{\today}

\begin{abstract}
Variational quantum algorithms dominate contemporary gate-based quantum enhanced optimisation, eigenvalue estimation and machine learning.  Here we establish the quantum computational universality of variational quantum computation by developing two objective functions which minimise to prepare outputs of arbitrary quantum circuits.   The fleeting resource of variational quantum computation is the number of expected values which must be iteratively minimised using classical-to-quantum outer loop optimisation.  An efficient solution to this optimisation problem is given by the quantum circuit being simulated itself.  The first construction is efficient in the number of expected values for $n$-qubit circuits containing $\mathcal{O}({poly} \ln n)$ non-Clifford gates---the number of expected values has no dependence on Clifford gates appearing in the simulated circuit.  The second approach yields $\mathcal{O}(L^2)$ expected values while introducing not more than $\mathcal{O}(\ln L)$ slack qubits, for a quantum circuit partitioned into $L$ gates.  Hence, the utilitarian variational quantum programming procedure---based on the classical evaluation of objective functions and iterated feedback---is in principle as powerful as any other model of quantum computation.  This result elevates the formal standing of the variational approach while establishing a new universal model of quantum computation.  
\end{abstract}

\maketitle

Variational quantum algorithms reduce quantum state preparation requirements while necessitating measurements of individual qubits in the computational basis \cite{2014NatCo...5E4213P, 2014arXiv1411.4028F}.   In the contemporary noisy intermediate-scale quantum (NISQ) enhanced technology setting \cite{2018arXiv180100862P}, a sought reduction in coherence time is mediated through an iterative classical-to-quantum feedback and optimization process.  Systematic errors which map to deterministic yet unknown control parameters---such as time-variability in the application of specific Hamiltonians or poor pulse timing---can have less impact on variational algorithms, as states are prepared iteratively and varied over to minimize objective function(s).  These experimental advantages have made the variational approach to quantum computation the most widely studied gate-based approach today. See the reviews \cite{Hadfield_2019, Moll_2018,Benedetti_2019}. 

A variational quantum algorithm executes to prepare a state which minimizes an objective function.  In the case of variational quantum approximate optimization (QAOA \cite{2014arXiv1411.4028F}), a state is prepared by alternating a Hamiltonian representing a penalty function (such as the \NP{}-hard Ising embedding of \SAT) with a Hamiltonian representing local tunneling terms.  The state is measured and the resulting bit string serves as a candidate solution to minimize the penalty function.  In the case of variational eigenvalue minimization (VQE \cite{2014NatCo...5E4213P}), the state is repeatedly prepared and measured to obtain a set of expected values which is term-wise calculated and collectively minimized.  In both approaches, the minimization process is iterated by updating Hamiltonian application times and/or gate angles.  Several techniques in variational quantum computation are closely related to methods appearing in machine learning \cite{2017arXiv171205304V, 2017Natur.549..195B} in which a quantum circuit is tuned subject to a given training dataset.  


Computational universality is a property of central study in both classical and quantum models of computation. Universal models such as adiabatic quantum computation~\cite{2004quant.ph..5098A} both discrete and continuous quantum walks~\cite{PhysRevLett.102.180501, Lovett_2010} and measurement based quantum computation~\cite{PhysRevLett.97.150504} have been proved computationally universal through constructions to emulate a universal set of quantum gates. This implies directly that the system has access to any polynomial time quantum algorithm on $n$-qubits (the power of quantum algorithms in the class \BQP{}). 

An alternative notion of {\it universality} also exists in the literature.  This strong notion is algebraic, wherein a system is called universal if its generating Lie algebra is proven to span {\bf su}$(2^n)$ for $n$ qubits.  We call this {\it controllability}.  

Evidently these two notions of universality can be interrelated: by proving that a controllable system can efficiently simulate a universal gate set, a controllable system becomes computationally universal.  It is conversely anticipated by the strong Church-Turing-Deutsch principle~\cite{1985RSPSA.400...97D}, that a computationally universal system can be made to simulate any controllable system. 

Herein we assume access to control sequences which can create quantum gates such as \cite{PhysRevLett.90.247901, PhysRevA.70.032314, 2002quant.ph..5115S}.  Given a quantum circuit of $L$ gates, preparing a state $\ket{\psi} = \prod_{l=1}^L U_l \ket{0}^{\otimes n}$ for unitary gates $U_l$, we construct a universal objective function that is minimised by $\ket{\psi}$.  The objective function is engineered to have certain desirable properties.  Importantly, we construct a gaped Hamiltonian where minimisation past some fixed tolerance ensures sufficient overlap with the desired output state $\ket{\psi}$.  

Recent work of interest by Lloyd considered the controllability of QAOA sequences to create quantum gates \cite{2018arXiv181211075L}.  The original goal of QAOA was to alternate target- and driver-Hamiltonians to evolve a system close to the target Hamiltonian's ground state---thereby solving an optimization problem instance. Lloyd showed that alternating driver and target Hamiltonians can be programmed to give computationally universal dynamics \cite{2018arXiv181211075L}.  Following Lloyd's work \cite{2018arXiv181211075L}, myself with two coauthors formulated Lloyd's QAOA controllability result past one-dimensional lines of qubits \cite{morales2019universality}. 

We assume controllable state preparation followed by local Pauli measurements.  These measurements are used to calculate an objective function (a Hamiltonian) which we prove minimizes to certify that the output of a target quantum circuit has been prepared. The objective function is expressed as a sum of terms in the Pauli basis. Each term corresponds to a simplistic measurement operator and hence this approach avoids the long sequences of gates used to simulate Hamiltonian's in the gate model.  While the objective function can be evaluated term-wise, achieving tolerance $\sim\epsilon$ requires $\sim\epsilon^{-2}$ measurements---see Hoeffding's inequality~\cite{doi:10.1080/01621459.1963.10500830}). 



{\bf Structure.} After introducing variational quantum computation as it applies to our setting, we construct an objective function (named a telescoping construction).  The number of expected values has no dependence on Clifford gates appearing in the simulated circuit and is efficient for circuits with $\mathcal{O}(\textup{poly} \ln n)$ non-Clifford gates, making it amenable for near term demonstrations.   We then modify the Feynman-Kitaev clock construction and prove that universal variational quantum computation is possible by minimising $\mathcal{O}(L^2)$ expected values while introducing not more than $\mathcal{O}(\ln L)$ slack qubits, for a quantum circuit partitioned into $L$ Hermitian blocks.  

We conclude by considering how the universal model of variational quantum computation can be utilised in practice.  In particular, the given gate sequence prepares a state which will minimise the objective function.  In practice, we think of this as providing a starting point for a classical optimizer. Given a $T$ gate sequence, we consider the first $L \leq T$ gates.  This $L$ gate circuit represents an optimal control problem where the starting point is the control sequence to prepare the $L$ gates.  The goal is to modify this control sequence (shorten it) using a variational feedback loop.  One would iterate this scenario, increasing $L$ up to $T$.  

\section{Variational Quantum Computation}

We work in the standard setting of quantum computation using $n$ qubits, where we typically fix the computational basis $\mathcal{B}^{\otimes n}$ and use the standard qubit representation of the Pauli group-algebra, satisfying the product identity 
$ X Y =  \imath Z$, together with $X^2 = Y^2 = Z^2 = \openone$.   
We will interchange the notation $\sigma_0 \equiv \openone$, $\sigma_1 \equiv X$, $\sigma_2 \equiv Y$, $\sigma_3 \equiv Z$ and consider Hamiltonian's $\mathcal{H} = \mathcal{H}^\dagger$ which act on the space of linear qubit maps $\mathcal{L}(\mathcal{B}^{\otimes n})$.  

We wish to simulate the output of an $L$ gate quantum circuit acting on the $n$-qubit product state $\ket{0}^{\otimes n}$.  We have access to $p$ appropriately bounded and tunable parameters to prepare and vary over a family of quantum states. All coefficients herein are assumed to be accurate to not more than $\textup{poly}(n)$ decimal places.  We will define an objective function that when minimized will produce a state close to the desired quantum circuit output. We will provide a solution to the minimization problem.  Establishing our results requires several definitions and supporting lemmas.  Proofs not appearing after a Lemma or Theorem can be found in the Supplementary Materials. 

\begin{definition}[Variational Statespace]
The variational statespace $\Gamma$ of a $p$-parameterized $n$-qubit state preparation process is the union of $\ket{\psi({\boldsymbol \theta})}$ over real assignments of ${\boldsymbol \theta}$, 
\begin{equation}
    \Gamma \bydef \bigcup_{{\boldsymbol \theta} \subset \mathbb{R}^{\times p}} \{ \ket{\psi({\boldsymbol \theta})} \} \subseteq \mathbb{C}_2^{\otimes n}. 
\end{equation} 
\end{definition} 

Variational statespace examples include preparing $\ket{\psi({\boldsymbol \theta})}$ by a fixed quantum circuit (called ansatz) with e.g.~${\boldsymbol \theta} \in [0, 2\pi)^{\times p}$ tunable parameters as 
\begin{equation}\label{eqn:cir}
\ket{\psi({\boldsymbol \theta})} = \Pi _{l=1}^LU_l \ket{0}^{\otimes n}, 
\end{equation}
where typically $U_l$ is adjusted by $\theta_l$ for $l=1$ to $p$, though some gates might be fixed. There are various approaches to implement ansatz circuits, including the Hardware Efficient Ansatz~\cite{2017Natur.549..242K} which implements a daisy chain of two-body coupling gates or the brick layer (a.k.a.~checkerboard) ansatz which is simply alternating stacks of nearest neighboring coupling gates (see for example~\cite{2020PhRvA.102a2415U}). 
 

  
\begin{definition}[Variational Sequence] \label{def:sequence}
A variational sequence is an assignment of ${\boldsymbol \theta}'$ to prepare a single state in a variational statespace $\ket{\psi({\boldsymbol \theta}')}\in \Gamma$.  
\end{definition}

Examples of variational sequences include the general expression of a fixed quantum gate sequence acting on a product state, which can be expressed as \eqref{eqn:cir}. 

We will define an objective function that can be efficiently calculated given access to a suitable quantum processor.  Minimization of this objective function will be proven to produce a close 2-norm approximation to the output of a given quantum circuit.  Let us consider the most suitably general Hamiltonian acting on qubits \eqref{eqn:Ham}. 


\begin{definition}[Objective Function] \label{def:ofunc}
We consider an objective function as the expected value of an operator expressed with real coefficients $\mathcal{J}^{a_1 a_2 \dots a_n}_{\alpha_1 \alpha_2 \dots \alpha_n}$ in the Pauli basis as, 
\begin{equation}\label{eqn:Ham}
  \mathcal{H} = \sum^{\textup{poly}(n)} \mathcal{J}^{a_1 a_2 \dots a_n}_{\alpha_1 \alpha_2 \dots \alpha_n} ~ \sigma^{a_1}_{\alpha_1} \sigma^{a_2}_{\alpha_2} \cdots \sigma^{a_n}_{\alpha_n} 
\end{equation}
where Greek letters index Pauli matrices, Roman letters index qubits and the sum is over a $\textup{poly}(n)$ bounded subset of the $4^n$ elements in the basis. The tensor ($\otimes$) is omitted in \eqref{eqn:Ham}. 
\end{definition}

We are concerned primarily with Hamiltonians where $\mathcal{J}^{a_1 a_2 \dots a_n}_{\alpha_1 \alpha_2 \dots \alpha_n}$ is given and known to be non-vanishing for at most some $\textup{poly}(n)$ terms.  This wide class includes Hamiltonians representing electronic structure \cite{2017Natur.549..242K}.  More generally, such Hamiltonians are of bounded cardinality. 

\begin{definition}[Cardinality] \label{def:cardinality}
The number of terms in the Pauli basis $\{\openone , X, Y, Z\}^{\otimes n}$ needed to express an objective function.  
\end{definition}

\begin{definition}[Bounded Objective Function] \label{def:efcom}
An instance is called {\it bounded} when it is taken from a uniform family of objective functions of cardinality bounded by $\textup{poly}(n)$. 
\end{definition}

We will now use the Cartesian tensor ($\times$) as states need not be proximally interacting and can be independently prepared.  Hence, $\ket{\phi}^{\times \mathcal{O}(\textup{poly}(n))}$ means we prepare polynomial many non-interacting copies of $\ket{\phi}$ to approximate each expected value.  

\begin{definition}\label{def:of}{\bf (Poly-Computable Objective Function)}
An objective function
\begin{equation}
    f:\ket{\phi}^{\times \mathcal{O}(\textup{poly}(n))} \rightarrow \mathbb{R}_{\geq 0}
\end{equation}
is called poly-computable provided $\textup{poly}(n)$ independent physical copies of $\ket{\phi}$ can be efficiently prepared to evaluate a bounded objective function.  
\end{definition}  

Efficiently computable objective function examples include calculating the expected value of $\mathcal{O}(\ln n)$ products and sums over
\begin{equation}
    \{\mathcal{H}, \langle \mathcal{H}\rangle, \cdot, +, \mathbb{R}\}
\end{equation}
for bounded cardinality $\mathcal{H}$. Examples include: 
\begin{enumerate}
\item[(2.i)] Calculating the expected value of $\mathcal{H}$ itself, which includes electronic structure Hamiltonians \cite{2014NatCo...5E4213P}. 
\item[(2.ii)] Calculating the dispersion var$(\mathcal{H}) = \langle \mathcal{H}^2\rangle- \langle \mathcal{H}\rangle^2$ which vanishes if and only if the prepared state is an eigenstate of $\mathcal{H}$ \cite{Kardashin2020}.   
\end{enumerate}

Acceptance (as follows) must be shown by providing a solution to the optimisation problem defined by the objective function. 

\begin{definition}[Accepting a Quantum State]
An objective function $f$ {\it accepts} $\ket{\phi}$ when given $\mathcal{O}(\textup{poly}~n)$ copies of $\ket{\phi}$, 
\begin{equation}
    f(\ket{\phi}^{\times \mathcal{O}(\textup{poly}(n)}) = f(\ket{\phi}, \ket{\phi}, \dots, \ket{\phi})< \Delta
\end{equation}
 evaluates strictly less than a chosen real parameter $\Delta > 0$.  
  \end{definition} 

The following theorem (\ref{thm:e2overlap}) applies rather generally to variational quantum algorithms that minimise energy by adjusting a variational state to cause an objective function to accept. Herein acceptance will imply the preparation of a quantum state, which begs to establish the following. 

\begin{lemma}[Variational Stability]\label{thm:e2overlap}
Let non-negative $\mathcal{H}= \mathcal{H}^\dagger\in \mathcal{L}(\mathbb{C}_d)$ have spectral gap $\Delta$ and non-degenerate ground eigenvector $\ket{\psi}$ of eigenvalue $0$.  
Consider then a unit vector $\ket{\phi}\in \mathbb{C}_d$ such that 
\begin{equation}
\bra{\phi }\mathcal{H}\ket{\phi } < \Delta 
\end{equation} 
it follows that 
\begin{equation}
1 - \frac{\bra{\phi}\mathcal{H}\ket{\phi} }{\Delta} \leq | \braket{\phi}{\psi}|^2 \leq 1 - \frac{\bra{\phi}\mathcal{H}\ket{\phi} }{\text{Tr}\{ \mathcal{H}\}}.
\end{equation} 
\end{lemma}

\subsection{Maximizing projection onto a circuit} 

We will now explicitly construct an elementary Hermitian penalty function that is non-negative, with a non-degenerate lowest ($0$) eigenstate---so as to apply Lemma \ref{thm:e2overlap}.  Minimisation of this penalty function prepares the output of a quantum circuit.  

\begin{theorem}[Telescoping Construction] \label{thm:tele}
Consider $\prod_l U_l \ket{0}^{\otimes n}$ an $L$-gate quantum circuit preparing state $\ket{\psi}$ on $n$-qubits and containing not more than $\mathcal{O}(\textup{poly}(\ln n))$ non-Clifford gates. Then there exists a Hamiltonian $\mathcal{H}\geq0$ on $n$-qubits with $\textup{poly}(L, n)$ cardinality, a $(L, n)$-independent gap $\Delta$ and non-degenerate ground eigenvector $\in\textup{span}\{\prod_l U_l \ket{0}^{\otimes n}$\}.  In particular, a variational sequence exists causing the Hamiltonian to accept $\ket{\phi}$ viz., $0\leq \bra{\phi}\mathcal{H}\ket{\phi}  < \Delta$ then Lemma \ref{thm:e2overlap} implies stability.  
\end{theorem} 

To prove Theorem \ref{thm:tele} we first show existence of the penalty function. Construct Hermitian $\mathcal{H} \in \mathcal{L}(\mathbb C_2^{\otimes n})$ with $\mathcal{H}\geq 0$ such that there exists a non-degenerate  $\ket{\psi}\in \mathbb C_2^{\otimes n}$  with the property that $\mathcal{H}\ket{\psi}=0$.   Define $P_\phi$ as a sum of projectors onto product states, i.e. 
\begin{equation}\label{eqn:proj}
P_\phi = \sum_{i=1}^n \ket{1}\bra{1}^{(i)} = \frac{n}{2}\left(\openone - \frac{1}{n}\sum_{i=1}^n   Z^{(i)} \right) 
\end{equation} 
and consider \eqref{eqn:proj} as the initial Hamiltonian, preparing state $\ket{0}^{\otimes n}$.

We will act on \eqref{eqn:proj} with a sequence of gates $ \prod_{l=1}^L U_l$ corresponding to the circuit being simulated as 
\begin{equation}\label{eqn:isoaffine}
h(k) = \left(\prod_{l=1}^{k\leq L} U_l \right)P_\phi \left(\prod_{l=1}^{k\leq L} U_l\right)^\dagger \geq 0
\end{equation}
which preserves the spectrum (i.e.~$P_\phi \ket{x} = |x|_1\ket{x}$ for $x\in \{0,1\}^n$ and $|\cdot|_1$ the Hamming weight). From the properties of $P_\phi$ it hence follows that $h(k)$ is non-negative and non-degenerate $\forall k \leq L$.  We now consider the action of the gates \eqref{eqn:isoaffine} on \eqref{eqn:proj}. 

At $k=0$ from \eqref{eqn:proj} there are $n$ expected values to be minimized plus a global energy shift that will play a multiplicative role as the circuit depth increases. To consider $k=1$ we first expand a universal gate set expressed in the linear extension of the Pauli basis.  

Interestingly, the coefficients $\mathcal{J}^{a b \dots c}_{\alpha \beta \dots \gamma}$ of the gates will not serve as direct input(s) to the quantum hardware; these coefficients play a direct role in the classical  step where the coefficients weight the sum to be minimized.  Let us then consider single qubit gates, in general form viz.,  
\begin{equation}
e^{-\imath {\bf{a}.\boldsymbol {\sigma}} \theta} = \openone \cos(\theta) - \imath {\bf{a}.\boldsymbol {\sigma}} \sin(\theta) 
\end{equation} 
where $\bf a$ is a unit vector and ${\bf{a}.\boldsymbol {\sigma}}  = \sum_{i=1}^3 a_i\sigma_i$.  So each single qubit gate increases the number of expected values by a factor of at most $4^2$.  At first glance, this appears prohibitive yet there are two factors to consider.  The first is the following Lemma (\ref{lemma:invariance}). 

\begin{lemma}[Clifford Gate Cardinality Invariance] \label{lemma:invariance} 
Let $\mathcal{C}$ be the set of all Clifford circuits on $n$ qubits, and let $\mathcal{P}$ be the set of all elements of the Pauli group on $n$ qubits.  Let $C\in\mathcal{C}$ and $P\in\mathcal{P}$ then it can be shown that $CPC^\dagger \in \mathcal{P}$ or in other words $C\left(\sigma^a_\alpha \sigma^b_\beta \cdots \sigma^c_\gamma \right)C^\dagger = \sigma^{a'}_{\alpha'} \sigma^{b'}_{\beta'} \cdots \sigma^{c'}_{\gamma'}$ and so Clifford circuits act by conjugation on tensor products of Pauli operators to produce tensor products of Pauli operators.   
\end{lemma}

For some $U$ a Clifford gate, Lemma \ref{lemma:invariance} shows that the cardinality of \eqref{eqn:isoaffine} is invariant.  Non-Clifford gates increase the cardinality by factors $\mathcal{O}(e^n)$ and so must be logarithmically bounded from above.  Hence, telescopes bound the number of expected values by restricting to circuit's with $k\sim\mathcal{O}(\textup{poly} \ln n)$ non-Clifford single qubit gates. Clifford gates do however modify the locality of terms appearing in the expected values---this is i.e.~prohibitive in adiabatic quantum computation yet arises here as local measurements. 

A final argument supporting the utility of telescopes is that the initial state is restricted primarily by the initial Hamiltonian having only a polynomial number of non-vanishing coefficients in the Pauli basis.  In practice---using today's hardware---it should be possible to prepare an $\epsilon$-close 2-norm approximation to any product state $\bigotimes_{k=1}^n \cos \theta_k \ket{0}+e^{\imath \phi_k}\sin \theta_k \ket{1}$ which is realised by modifying the projectors in \eqref{eqn:proj} with a product of single qubit maps $\bigotimes_{k=1}^n U_k$.  Other more complicated states would also be possible.  

To finish the proof of Lemma \ref{thm:tele}, the variational sequence is given by the description of the gate sequence itself.  That is, 
\begin{equation}
    h(k) \left(\prod_{l=1}^{k\leq L} U_l \right) \ket{x} = |x|_1\left(\prod_{l=1}^{k\leq L} U_l \right) \ket{x}
\end{equation}
is minimized for $x$ the string of zeros in $\{0,1\}^n$. Hence a state can be prepared causing the Hamiltonian to accept and stability applies (Lemma \ref{thm:e2overlap}). 

To explore telescopes in practice, let us then explicitly consider the quantum algorithm for state overlap (a.k.a., {\it swap test} see e.g.~\cite{2018NJPh...20k3022C}).  This algorithm has an analogous structure to phase estimation, a universal quantum primitive of error-corrected quantum algorithms.  

\begin{example} 
We are given two $d$-qubit states $\ket{\rho}$ and $\ket{\tau}$ which will be non-degenerate and minimal eigenvalue states of some initial Hamiltonian(s) on $n+1$ qubits 
\begin{equation}
h(0) \ket{+, \rho, \tau} = 0 
\end{equation}
corresponding to the minimization of $\textup{poly}(n/2)+1$ expected values where the first qubit (superscript 1 below) adds one term and is measured in the $X$-basis. The controlled swap gate takes the form 
\begin{equation}
[U_\text{swap}]^1_m=\frac{1}{2}\left(\openone^1 +Z^1\right)\otimes \openone^m + \frac{1}{2}\left(\openone^1 - Z^1\right)\otimes \mathcal{S}^m
\end{equation} 
where $m=(i,j)$ indexes a qubit pair and the exchange operator of a pair of qubit states is $\mathcal{S}=\openone + \boldsymbol {\sigma}.\boldsymbol {\sigma}$.  For the case of $d=1$ we arrive at the simplest (3-qubit) experimental demonstration.  At the minimum ($=0$), the expected value of the first qubit being in logical zero is $\frac{1}{2}+\frac{1}{2}|\braket{\rho}{\tau}|^2$.  The final Hadamard gate on the control qubit is considered in the measurement step. 
\end{example} 

Telescopes provide some handle on what we can do without adding additional slack qubits yet fail to directly prove universality in their own right.  The crux lies in the fact that we are only allowed some polynomial in $\ln n$ non-Clifford gates (which opens an avenue for classical simulation, see \cite{Bravyi_2016, Bravyi_2019}).  Interestingly however, we considered the initial Hamiltonian in \eqref{eqn:proj} as a specific sum over projectors.  We instead could bound the cardinality by some polynomial in $n$.  Such a construction will now be established: requires the addition of slack qubits. The universal construction then follows. 

\subsection{Maximizing projection onto the history state}

We will now prove the following theorem (\ref{thm:history}) which establishes universality of the variational model of quantum computation. 

\begin{theorem}\label{thm:history}
Consider a quantum circuit of $L$ gates on $n$-qubits producing state $\prod_l U_l \ket{0}^{\otimes n}$.  Then there exists an objective function (Hamiltonian, $\mathcal{H}$) with non-degenerate ground state, cardinality $\mathcal{O}(L^2)$ and spectral gap $\Delta\geq \mathcal{O}(L^{-2})$ acting on $n+\mathcal{O}(\ln L)$ qubits such that acceptance implies efficient preparation of the state $\prod_l U_l \ket{0}^{\otimes n}$.  Moreover, a variational sequence exists causing the objective function to accept.  
\end{theorem} 

To construct an objective function satisfying Theorem \ref{thm:history}, we modify the Feynman-Kitaev clock construction \cite{Fey82, KSV02}.  Coincidentally (and tangential to our objectives here), this construction is also used in certain definitions of the complexity class quantum-Merlin-Arthur (\QMA{}), the quantum analog of \NP{}, through the \QMA{}-complete problem k-\LH{}~\cite{KSV02}.  

Feynman developed a time-independent Hamiltonian that induces unitary dynamics to simulate a sequence of gates \cite{Fey82}. Consider Feynman's Hamiltonians:
\begin{equation} \label{eqn:hpropfey}
\begin{split}
\tilde{\mathcal{H}}_t &= U_t\otimes \ket{t}\bra{t-1} + U_t^\dagger \otimes \ket{t-1}\bra{t} \\
 \tilde{\mathcal{H}}_{\text{prop}} & = \sum_{t=1}^L \tilde{\mathcal{H}}_t
 \end{split}
\end{equation}
where the Hamiltonian \eqref{eqn:hpropfey} acts on a clock register (right of $\otimes$) with orthogonal clock states $0$ to $L$ and an initial state $\ket{\xi}$ (left).  Observation of the clock in state $\ket{L}$ after some time $s=s_\star$ produces 
\begin{equation}
\openone \otimes \bra{L} e^{-\imath \cdot s \cdot \mathcal{H}_{\text{prop}}} \ket{\xi}\otimes \ket{0} = U_L \cdots U_1 \ket{\xi}. 
\end{equation} 

The Hamiltonian $\mathcal{H}_{\text{prop}}$ in \eqref{eqn:hpropfey} can be modified as \eqref{eqn:hprop2} so as to have the history state \eqref{eqn:hist} as its ground state 
\begin{align} \label{eqn:hprop2}
&- U_t\otimes \ket{t}\bra{t-1} - U_t^\dagger \otimes \ket{t-1}\bra{t} + \ket{t}\bra{t} + \ket{t-1}\bra{t-1}\nonumber \\
 & = 2\cdot \mathcal{H}_t \geq 0 
\end{align}
where $\mathcal{H}_t$ is a projector.  Then $\mathcal{H}_{\text{prop}} = \sum_{t=1}^L \mathcal{H}_t$ has the history state  
\begin{equation}\label{eqn:hist}
\ket{\psi_{\text{hist}}} = \frac{1}{\sqrt{L+1}} \sum_{t=0}^L U_t\cdots U_1\ket{\xi}\otimes \ket{t} 
\end{equation}
as its ground state as for any input state $\ket{\xi}$ where $0 = \bra{\psi_{\text{hist}}}\mathcal{H}_{\text{prop}} \ket{\psi_{\text{hist}}}$.  This forms the building blocks of our objective function.  We will hence establish Theorem \ref{thm:history} by a series of lemma.  We let $P_0 \bydef \ket{0}\bra{0}$. 

\begin{lemma}[Degeneracy Lifting] \label{lemma:degen}
Adding the tensor product of a projector on the first clock qubit with a telescope  
\begin{equation}
\mathcal{H}_{\text{in}} = V\left( \sum_{i = 1}^n P_1^{(i)} \right)V^\dagger \otimes P_0
\end{equation} 
lifts the degeneracy of the ground space of $\mathcal{H}_{\text{prop}}$ and the history state with fixed input as 
\begin{equation}
\frac{1}{\sqrt{L+1}} \sum_{t=0}^L \prod_{l = 1}^t U_l(V\ket{0}^{\otimes n}) \otimes \ket{t} 
\end{equation}
is the non-degenerate ground state of $J\cdot \mathcal{H}_{\text{in}} + K\cdot  \mathcal{H}_{\text{prop}}$ for real $J, K>0$.  
\end{lemma} 

\begin{lemma}[Gap Existence]\label{lemma:gap}
For appropriate non-negative $J$ and $K$, the operator $J\cdot \mathcal{H}_{\text{in}} + K\cdot  \mathcal{H}_{\text{prop}}$ is gapped with a non-degenerate ground state and hence, Lemma \ref{thm:e2overlap} applies with 
\begin{equation}
\Delta \geq \max\{ J,  \frac{K \pi^2}{2(L+1)^2} \}. 
\end{equation} 
\end{lemma} 

\begin{lemma}[Logspace Embedding $\mathcal{H}_{\text{prop}}$] \label{lemma:logem}
The clock space of $\mathcal{H}_{\text{prop}}$ embeds into $\mathcal{O}(\ln L)$ slack qubits, leaving the ground space of $J\cdot \mathcal{H}_{\text{in}} + K\cdot  \mathcal{H}_{\text{prop}}$ and the gap invariant. 
\end{lemma}

\begin{lemma}[Existence and Acceptance] \label{lemma:uob}
The objective function $J\cdot \mathcal{H}_{\text{in}} + K\cdot  \mathcal{H}_{\text{prop}}$ satisfies Theorem \ref{thm:history}. 
The gate sequence $\prod_l U_l \ket{0}^{\otimes n}$ is accepted by the objective function from Lemma \ref{lemma:uob} thereby satisfying Theorem \ref{thm:history}. 
\end{lemma}

We will add $K$ identity gates to boost the probability of the desired circuit output state $\ket{\phi } = \Pi _{l=1}^LU_l \ket{0}^{\otimes n}$. From Lemma \ref{thm:e2overlap}, we have that 
\begin{equation}\label{eqn:application}
1 - \frac{\bra{\phi}\mathcal{H}\ket{\phi} }{\Delta} \leq | \braket{\phi}{\psi_{\text{hist}}}|^2 = \frac{1}{1+\frac{L+1}{K}}
\end{equation} 
whenever $\bra{\phi }\mathcal{H}\ket{\phi } < \max\{ J,  \frac{K \pi^2}{2(L+1)^2} \}$.  For large enough $K>L$, the right hand side of \eqref{eqn:application} approaches unity, satisfying the theorem. 

Finally are faced with considering self-inverse gates.  Such gates ($U$) have a spectrum $\text{Spec}(U)\subseteq\{\pm1\}$, are bijective to idempotent projectors ($P^2=P=P^\dagger$), viz.~$U = \openone - 2P$ and if $V$ is a self-inverse quantum gate, so is the unitary conjugate $\tilde{V}= G V G^\dagger$ under arbitrary $G$.  Shi showed that a set comprising the controlled not gate (a.k.a.~Feynman gate) plus any one-qubit gate whose square does not preserve the computational basis is universal \cite{2002quant.ph..5115S}.  Consider Hermitian 
\begin{equation}
R(\theta) = X\cdot \sin(\theta) + Z \cdot \cos(\theta), 
\end{equation} 
then 
\begin{equation}\label{eqn:RR}
e^{\imath \theta Y} = R(\pi / 2) \cdot R(\theta). 
\end{equation} 
Hence, a unitary $Y$ rotation is recovered by a product of two Hermitian operators.  A unitary $X$ rotation is likewise recovered by the composition \eqref{eqn:RR} when considering Hermitian $Y\cdot \sin(\theta) - Z \cdot \cos(\theta)$.  The universality of self-inverse gates is then established, with constant overhead.  Hence and to conclude, the method introduces not more than $\mathcal{O}(L^2)$ expected values while requiring not more than $\mathcal{O}(\ln L)$ slack qubits, for an $L$ gate quantum circuit. 

\subsection{Ansatz states and a combinatorial quantum circuit area law} 

Our construction of universal variational quantum computation has not considered whether a restricted form of ansatz is capable of universal quantum computation at some arbitrary depth as Lloyd~\cite{2018arXiv181211075L} and others~\cite{morales2019universality} have.  Instead, the objective function to be minimised is defined in terms of the unitary gates arising in the target circuit to be simulated.  What ansatz states are then required to simulate a given target circuit?

This question appears to be difficult and not much is currently known.  In the case of QAOA it was recently shown by myself and coauthors that the ability of an ansatz to approximate the ground state energy of a satisfiability instance worsens with increasing problem density (the ratio of constraints to variables) \cite{2019arXiv190611259A}. These related results however do not imediately apply to our interests here.  

Towards our goals, we show that reasonable depth circuits might saturate bipartite entanglement---the depth of these circuits scales with the number of qubits and also depends on the interaction geometry present in a given quantum processor.  Consider the following.  

An ebit is a unit of entanglement contained in a maximally entangled two-qubit (Bell) state.  A quantum state with $q$ ebits of entanglement (quantified by any entanglement measure) contains the same amount of entanglement (in that measure) as $q$ Bell states.  
\begin{lemma}
 Let $c$ be the depth of 2-qubit controlled rotation gates in the $n$-qubit hardware-efficient ansatz.  Then the maximum possible number of ebits across any bipartition is 
    $$ E_b = \min \{ \left \lfloor{n/2}\right \rfloor, c \}$$
 \end{lemma} 
In a low-depth circuit, the underlying geometry of the processor heavily dictates $c$ above.  For example, for a line of qubits and for a ring, the minimal $c$ required to possibly maximise $E_b$ is $\sim n/2$ and $\sim n/4$ respectfully. 
However, in the case of a grid, the minimal depth scales as $\sim \sqrt{2}/2$.  

Hence, if we wish to simulate a quantum algorithm described by a low-depth circuit, having access to a grid architecture could provide an intrinsic advantage.  Specifically, our combinatorial quantum circuit area law establishes that  an objective circuit generating $k< \left \lfloor{n/2}\right \rfloor$ ebits across every bipartition, must be simulated by an ansatz of at least minimal required circuit depth $\sim \sqrt{k}$ on a grid.  

While this does establish a preliminary relationship, the general case remains unclear at the time of writing.  For example, given a quantum circuit with application time $t^\star$ which outputs $\ket{\psi}$, what is the minimal $t(\epsilon)\leq t^\star$ for a control sequence to provide an $\epsilon$ close 2-norm approximation to $\ket{\psi}$?

\section{Discussion} 

We have established that variational quantum computation admits a universal model.  The gate sequence being simulated serves as an upper-bound showing that a control sequence exists to minimize the expected values.  Expected values are then the fleeting resource of the variational model. 

Although error correction is assumed in our universality proofs, the techniques we develop should augment possibilities in the NISQ setting, particularly with the advent of error suppression techniques  \cite{2016NJPh...18b3023M, PhysRevX.7.021050}.  Importantly, variational quantum computation forms a universal model in its own right and is not (in principle) limited in application scope.  

An interesting feature of variational quantum computation is how many-body Hamiltonian terms are realized as part of the measurement process.  This is in contrast with leading alternative models of universal quantum computation.  

In the gate model, many-body interactions must be simulated by sequences of two-body gates.  The adiabatic model applies perturbative gadgets to approximate many-body interactions with two-body interactions  \cite{2004quant.ph..5098A, BL08}.    The variational model simulates many body interactions by local measurements.  Moreover the coefficients weighting many-body terms need not be implemented by the quantum hardware directly; this weight is compensated for in the classical process.   Finally, as many quantum states can cause a considered objective function to accept, the presented model is therefore partially agnostic to how states are prepared. 
 
Variational counter parts to an increasing number of celebrated quantum algorithms have been recently developed, including the solution to linear~\cite{bravoprieto2020variational} (and non-linear~\cite{PhysRevA.101.010301}) systems.  Indeed, computational universality implies that such variational incarnations generally exist.  Yet the present results certainly don't rule out significant overhead reductions of task tailored variaitonal quantum algorithms.

\begin{acknowledgments}
The author acknowledges support from the project, {\it Leading Research Center on Quantum Computing} (Agreement No.~014/20). 
{\it Competing interests.} The author declare no competing interests.
{\it Data and code availability.} The data that supports the findings of this study are available within the article. 
\end{acknowledgments}

\subsection*{Methods} 
\noindent Data Availability. No data sets were generated.  \\

\noindent Materials supporting the conclusions drawn in this study are available in the supplementary information.

\onecolumngrid
\bibliography{pra-sample.bib}

\begin{thebibliography}{10}

\bibitem{2004quant.ph..5098A}
D.~{Aharonov}, W.~{van Dam}, J.~{Kempe}, Z.~{Landau}, S.~{Lloyd}, and
  O.~{Regev}.
\newblock Adiabatic quantum computation is equivalent to standard quantum
  computation.
\newblock In {\em 45th Annual IEEE Symposium on Foundations of Computer
  Science}, pages 42--51, 2004.

\bibitem{2019arXiv190611259A}
V.~{Akshay}, H.~{Philathong}, M.~E.~S. {Morales}, and J.~{Biamonte}.
\newblock Reachability deficits in quantum approximate optimization.
\newblock {\em Physical Review Letters}, 124(9), Mar 2020.

\bibitem{Benedetti_2019}
Marcello Benedetti, Erika Lloyd, Stefan Sack, and Mattia Fiorentini.
\newblock Parameterized quantum circuits as machine learning models.
\newblock {\em Quantum Science and Technology}, 4(4):043001, Nov 2019.

\bibitem{PhysRevLett.90.247901}
Simon~C. Benjamin and Sougato Bose.
\newblock Quantum computing with an always-on heisenberg interaction.
\newblock {\em Phys. Rev. Lett.}, 90:247901, Jun 2003.

\bibitem{PhysRevA.70.032314}
Simon~C. Benjamin and Sougato Bose.
\newblock Quantum computing in arrays coupled by ``always-on'' interactions.
\newblock {\em Phys. Rev. A}, 70:032314, Sep 2004.

\bibitem{2017Natur.549..195B}
J.~{Biamonte}, P.~{Wittek}, N.~{Pancotti}, P.~{Rebentrost}, N.~{Wiebe}, and
  S.~{Lloyd}.
\newblock {Quantum machine learning}.
\newblock {\em Nature}, 549:195--202, September 2017.

\bibitem{BL08}
Jacob~D. {Biamonte} and Peter~J. {Love}.
\newblock {Realizable Hamiltonians for universal adiabatic quantum computers}.
\newblock {\em Physical Review A}, 78:012352, July 2008.

\bibitem{bravoprieto2020variational}
Carlos Bravo-Prieto, Ryan LaRose, M.~Cerezo, Yigit Subasi, Lukasz Cincio, and
  Patrick~J. Coles.
\newblock Variational quantum linear solver, 2020.

\bibitem{Bravyi_2019}
Sergey Bravyi, Dan Browne, Padraic Calpin, Earl Campbell, David Gosset, and
  Mark Howard.
\newblock Simulation of quantum circuits by low-rank stabilizer decompositions.
\newblock {\em Quantum}, 3:181, Sep 2019.

\bibitem{Bravyi_2016}
Sergey Bravyi and David Gosset.
\newblock Improved classical simulation of quantum circuits dominated by
  clifford gates.
\newblock {\em Physical Review Letters}, 116(25), Jun 2016.

\bibitem{PhysRevLett.102.180501}
Andrew~M. Childs.
\newblock Universal computation by quantum walk.
\newblock {\em Phys. Rev. Lett.}, 102:180501, May 2009.

\bibitem{2018NJPh...20k3022C}
Lukasz {Cincio}, Yi{\u{g}}it {Subaș{\i}}, Andrew~T. {Sornborger}, and
  Patrick~J. {Coles}.
\newblock {Learning the quantum algorithm for state overlap}.
\newblock {\em New Journal of Physics}, 20:113022, November 2018.

\bibitem{1985RSPSA.400...97D}
D.~{Deutsch}.
\newblock {Quantum theory, the Church-Turing principle and the universal
  quantum computer}.
\newblock {\em Proceedings of the Royal Society of London Series A},
  400(1818):97--117, Jul 1985.

\bibitem{2014arXiv1411.4028F}
Edward {Farhi}, Jeffrey {Goldstone}, and Sam {Gutmann}.
\newblock {A Quantum Approximate Optimization Algorithm}.
\newblock {\em arXiv e-prints}, page arXiv:1411.4028, November 2014.

\bibitem{Fey82}
Richard~P. Feynman.
\newblock Quantum mechanical computers.
\newblock {\em Optics News}, 11(2):11--20, Feb 1985.

\bibitem{Hadfield_2019}
Stuart Hadfield, Zhihui Wang, Bryan O’Gorman, Eleanor Rieffel, Davide
  Venturelli, and Rupak Biswas.
\newblock From the quantum approximate optimization algorithm to a quantum
  alternating operator ansatz.
\newblock {\em Algorithms}, 12(2):34, Feb 2019.

\bibitem{doi:10.1080/01621459.1963.10500830}
Wassily Hoeffding.
\newblock Probability inequalities for sums of bounded random variables.
\newblock {\em Journal of the American Statistical Association},
  58(301):13--30, 1963.

\bibitem{2017Natur.549..242K}
Abhinav {Kandala}, Antonio {Mezzacapo}, Kristan {Temme}, Maika {Takita}, Markus
  {Brink}, Jerry~M. {Chow}, and Jay~M. {Gambetta}.
\newblock {Hardware-efficient variational quantum eigensolver for small
  molecules and quantum magnets}.
\newblock {\em \nat}, 549:242--246, September 2017.

\bibitem{Kardashin2020}
Andrey Kardashin, Alexey Uvarov, Dmitry Yudin, and Jacob Biamonte.
\newblock Certified variational quantum algorithms for eigenstate preparation.
\newblock {\em Physical Review A}, 102(5), November 2020.

\bibitem{KSV02}
A.~Yu. Kitaev, A.~H. Shen, and M.~N. Vyalyi.
\newblock {\em Classical and Quantum Computation}.
\newblock American Mathematical Society, Boston, MA, USA, 2002.

\bibitem{PhysRevX.7.021050}
Ying Li and Simon~C. Benjamin.
\newblock Efficient variational quantum simulator incorporating active error
  minimization.
\newblock {\em Phys. Rev. X}, 7:021050, Jun 2017.

\bibitem{2018arXiv181211075L}
Seth {Lloyd}.
\newblock {Quantum approximate optimization is computationally universal}.
\newblock {\em arXiv e-prints}, page arXiv:1812.11075, December 2018.

\bibitem{Lovett_2010}
Neil~B. Lovett, Sally Cooper, Matthew Everitt, Matthew Trevers, and Viv Kendon.
\newblock Universal quantum computation using the discrete-time quantum walk.
\newblock {\em Physical Review A}, 81(4), Apr 2010.

\bibitem{PhysRevA.101.010301}
Michael Lubasch, Jaewoo Joo, Pierre Moinier, Martin Kiffner, and Dieter Jaksch.
\newblock Variational quantum algorithms for nonlinear problems.
\newblock {\em Phys. Rev. A}, 101:010301, Jan 2020.

\bibitem{2016NJPh...18b3023M}
Jarrod~R. {McClean}, Jonathan {Romero}, Ryan {Babbush}, and Al{\'a}n
  {Aspuru-Guzik}.
\newblock {The theory of variational hybrid quantum-classical algorithms}.
\newblock {\em New Journal of Physics}, 18:023023, February 2016.

\bibitem{Moll_2018}
Nikolaj Moll, Panagiotis Barkoutsos, Lev~S Bishop, Jerry~M Chow, Andrew Cross,
  Daniel~J Egger, Stefan Filipp, Andreas Fuhrer, Jay~M Gambetta, Marc Ganzhorn,
  and et~al.
\newblock Quantum optimization using variational algorithms on near-term
  quantum devices.
\newblock {\em Quantum Science and Technology}, 3(3):030503, Jun 2018.

\bibitem{morales2019universality}
M.~E.~S. Morales, J.~D. Biamonte, and Z.~Zimbor{\'{a}}s.
\newblock On the universality of the quantum approximate optimization
  algorithm.
\newblock {\em Quantum Information Processing}, 19(9), August 2020.

\bibitem{2014NatCo...5E4213P}
A.~{Peruzzo}, J.~{McClean}, P.~{Shadbolt}, M.-H. {Yung}, X.-Q. {Zhou}, P.~J.
  {Love}, A.~{Aspuru-Guzik}, and J.~L. {O'Brien}.
\newblock {A variational eigenvalue solver on a photonic quantum processor}.
\newblock {\em Nature Communications}, 5:4213, July 2014.

\bibitem{2018arXiv180100862P}
John {Preskill}.
\newblock {Quantum Computing in the NISQ era and beyond}.
\newblock {\em arXiv e-prints}, page arXiv:1801.00862, Jan 2018.

\bibitem{2002quant.ph..5115S}
Yaoyun Shi.
\newblock Both toffoli and controlled-not need little help to do universal
  quantum computing.
\newblock {\em Quantum Info. Comput.}, 3(1):84--92, January 2003.

\bibitem{2020PhRvA.102a2415U}
A.~V. {Uvarov}, A.~S. {Kardashin}, and J.~D. {Biamonte}.
\newblock {Machine learning phase transitions with a quantum processor}.
\newblock {\em \pra}, 102(1):012415, July 2020.

\bibitem{PhysRevLett.97.150504}
Maarten Van~den Nest, Akimasa Miyake, Wolfgang D\"ur, and Hans~J. Briegel.
\newblock Universal resources for measurement-based quantum computation.
\newblock {\em Phys. Rev. Lett.}, 97:150504, Oct 2006.

\bibitem{2017arXiv171205304V}
Guillaume {Verdon}, Michael {Broughton}, and Jacob {Biamonte}.
\newblock {A quantum algorithm to train neural networks using low-depth
  circuits}.
\newblock {\em arXiv e-prints}, page arXiv:1712.05304, December 2017.

\end{thebibliography}
\bibliographystyle{plain}

\clearpage
\onecolumngrid
\begin{center}
\textbf{\large Supplementary Materials}
\end{center}
\appendix
\begin{proof}[Lemma \ref{thm:e2overlap}]
Let $\{\ket{\psi_x}\}_{x=0}^{d-1}$ be the eigenbasis of $\mathcal{H}$.  The integer variable $x$ monotonically orders this basis by corresponding eigenvalue ($\lambda_0 = 0 \leq \lambda_1 = \Delta \leq\dots$). Consider $\ket{\phi}= \sum c_x \ket{\psi_x}$ and $P(x) = |c_x|^2$ with $c_x = \braket{\psi_x}{\phi}$.  Then 
\begin{equation}
 \bra{\phi}\mathcal{H}\ket{\phi} = \sum P(x)\lambda(x) = \sum_{\neg 0} P(x)\lambda(x)
\end{equation} 
Consider $1 = \braket{\phi}{\phi} = \sum P(x) = P(0) + \sum_{\neg 0} P(x)$.  So $1 - P(0) \geq P(x)~~ \forall x \neq 0$. Hence 
\begin{equation}
\sum_{\neg 0}P(x)\lambda(x) \leq (1-P(0))  \sum_{\neg 0}\lambda(x) 
\end{equation} 
From $\lambda(0) = 0$ we have that 
\begin{equation}
(1-P(0)) \sum\lambda(x) = (1-P(0))\cdot \text{Tr}\{\mathcal{H}\} = (1-|\braket{\psi_0}{\phi}|^2)\cdot \text{Tr}\{\mathcal{H}\}
\end{equation} 
from which the upper bound follows. 

For the lower bound, starting from the assumption $\bra{\phi}\mathcal{H}\ket{\phi} < \Delta$ we consider a function of the integers $\tilde{\lambda}(x)\cdot \Delta = \lambda(x)$ $\forall x>0$.  Then 
\begin{equation}
\bra{\phi}\mathcal{H}\ket{\phi} = \sum P(x)\lambda(x) = \sum_{\neg 0} P(x)\lambda(x) =  \Delta\cdot \sum_{\neg 0} P(x)\tilde{\lambda}(x)
\end{equation} 
hence $\sum_{\neg 0} P(x)\tilde{\lambda}(x) < 1$ as $\tilde{\lambda}(x)\geq 1$, $\forall x \geq 0$. We then have that 
\begin{equation}
\sum_{\neg 0} P(x) \leq \sum_{\neg 0} P(x)\tilde{\lambda}(x) < 1 
\end{equation} 
and so $1 - \sum_{\neg 0} P(x)\tilde{\lambda}(x)  \geq 0$.  Hence, 
\begin{equation}
1 = \braket{\phi}{\phi} = P(0) + \sum_{\neg 0} P(x) \leq P(0) + \sum_{\neg 0} P(x)\tilde{\lambda}(x) 
\end{equation} 
We establish that 
\begin{equation} 
1 \leq P(0) + \sum_{\neg 0} \frac{P(x)\lambda(x)}{\Delta} = |\braket{\psi_0}{\phi}|^2 + \frac{\bra{\phi}\mathcal{H}\ket{\phi}}{\Delta}
\end{equation} 
which leads directly to the desired lower bound.  \hfill $\blacksquare$
\end{proof} 

\begin{proof}[Lemma \ref{lemma:degen}]
The lowest energy subspace of $\mathcal{H}_{\text{prop}}$ is spanned by $\ket{\psi_{\text{hist}}}$ which has degeneracy given by freedom in choosing any input state $\ket{\xi}$.  To fix the input, consider a tensor product with a telescope 
\begin{equation}
\mathcal{H}_{\text{in}} = V\left( \sum_{i = 1}^n P_1^{(i)} \right)V^\dagger \otimes P_0
\end{equation} 
for $P_1 = \ket{1}\bra{1} = \openone - P_0$ acting on the qubit labeled in the subscript $(i)$ on the right hand side and on the clock space (left).  It is readily seen that $\mathcal{H}_{\text{in}}$ has unit gap and 
\begin{equation} 
\ker \{\mathcal{H}_{\text{in}}\} = \textup{span}\{ \ket{\zeta}\otimes \ket{c}, V\ket{0}^{\otimes n}\otimes \ket{0} | 0<c \in \mathbb{N}_+ \leq L, \ket{\zeta}\in \mathbb{C}_2^{\otimes n}\} 
\end{equation} 
Now for positive $J$, $K$ 
\begin{equation}
\arg\min \{J\cdot \mathcal{H}_{\text{in}} + K\cdot  \mathcal{H}_{\text{prop}} \} \propto \frac{1}{\sqrt{L+1}} \sum_{t=0}^L \prod_{l = 1}^t U_l(V\ket{0}^{\otimes n}) \otimes \ket{t} 
\end{equation} \hfill $\blacksquare$
\end{proof}

\begin{proof}[Lemma \ref{lemma:gap}] 
$\mathcal{H}_{\text{prop}}$ is diagonalized by the following unitary transform 
\begin{equation} 
W = \sum_{t=0}^L U_t\cdots U_1\otimes \ket{t}\bra{t} 
\end{equation} 
then $W\mathcal{H}_{\text{prop}}W^\dagger$ acts as identity on the register space (left) and induces a quantum walk on a 1D line on the clock space (right). Hence the eigenvalues are known to be $\lambda_k = 1-\cos\left(\frac{\pi k}{1+L}\right)$ for integer $0\leq k \leq L$. 
From the standard inequality, $1-\cos(x)\leq x^2/2$, we find that $\mathcal{H}_{\text{prop}}$ has a gap lower bounded as 
\begin{equation} 
\lambda_0 = 0 \leq \frac{\pi^2}{2(L+1)^2} \leq \lambda_1
\end{equation}  

From Weyl's inequalities, it follow that $J\cdot \mathcal{H}_{\text{in}} + K\cdot  \mathcal{H}_{\text{prop}}$ is gapped as
\begin{eqnarray} 
\lambda_0 = 0 &<& \max\{ \lambda_1(J\cdot \mathcal{H}_{\text{in}}), \lambda_1(K\cdot  \mathcal{H}_{\text{prop}}) \}\\
&\leq& \lambda_1(J\cdot \mathcal{H}_{\text{in}} + K\cdot  \mathcal{H}_{\text{prop}}) \\
&\leq& \min\{ \lambda_{n-1}(J\cdot \mathcal{H}_{\text{in}}), \lambda_{n-1}(K\cdot  \mathcal{H}_{\text{prop}}) \}
\end{eqnarray} 
with a non-degenerate ground state and hence, Lemma \ref{thm:e2overlap} applies with 
\begin{equation}
\Delta \geq \max\{ J,  \frac{K \pi^2}{2(L+1)^2} \}
\end{equation} \hfill $\blacksquare$
\end{proof} 

\begin{proof}[Lemma \ref{lemma:logem}] 
An $L$-gate circuit requires at most $k = \lceil \ln_2 L \rceil$ clock qubits. Consider a projector $P$ onto the orthogonal compliment of a basis state given by bit string ${\bf x} = x_1 x_2 \dots x_k$.  Then 
\begin{equation} \label{eqn:clockpro}
P_{\bf x} = \ket{\bar{\bf x}}\bra{\bar{\bf x}} = \bigotimes_{i=1}^{\lceil \ln_2 L \rceil} \frac{1}{2}\left(\openone + (-1)^{x_i} Z_i\right) 
\end{equation} 
where $\bar{\bf x}$ is the bitwise logical compliment of ${\bf x}$.  \hfill $\blacksquare$
\end{proof} 

\begin{proof}[Lemma \ref{lemma:uob}] Sketch. This term \eqref{eqn:clockpro} contributes $L$ terms and hence so does each of the four terms in $\mathcal{H}_t$ from \eqref{eqn:hprop2}.  Hence, the entire sum contributes $3\cdot L^2$ expected values, where we assume $U=U^\dagger$ and that $L$ is upper bounded by some family of circuits requiring ${\mathcal O}(\textup{poly}~ n)$ gates.  The input penalty $\mathcal{H}_{\text{in}}$ contributes $n$ terms and for an $L$-gate circuit on $n$-qubits we arrive at a total of $\mathcal{O}(\textup{poly}~ L^2)$ expected values and $\mathcal{O}(\lceil \ln_2 L \rceil)$ slack qubits.  Adding identity gates to the circuit can boost output probabilities, causing the objective function to accept for a state prepared by the given quantum circuit. \hfill $\blacksquare$
\end{proof} 

\section*{The hardware efficient ansatz satisfies a combinatorial quantum circuit area law}

Here we will consider some properties of the quantum states that are accessible in NISQ era quantum information processing.  Here we provide a bound for the minimal depth circuit (generated from the so called, hardware efficient Ansatz as used in recent experiments \cite{2017Natur.549..242K}) to possibly saturate bipartite entanglement on any bipartition.  Understanding the computational power of these circuits represents a central open question in the field of quantum computation today.  This appendix seeks to quantify contemporary capacities.  We are currently not able to express the success probability as a function of the circuit depth required for an objective function to accept. 

\begin{definition}[Interaction graph]    
Consider the Hamiltonian 
$$ \mathcal{H} = \sum_{ij} J_{ij} A_i A_j + \sum_i b_i B_i  + \sum_i c_i C_i $$
The support matrix $S$ of $J_{ij}$ is defined to have the entries
 $$ S_{ij} =  [J_{ij}]^0 $$
 and is called the \textit{interaction graph of $\mathcal{H}$}---a symmetric adjacency matrix.
\end{definition}

\begin{remark} 
The interaction graph induces a space-time quantum circuit defined by a \textit{tiling} on the multiplex from $S$.
\end{remark}

\begin{definition} 
 A \textit{Tiling} is a gate sequence acting on a multiplex network induced by $S$.  
 \end{definition} 
 
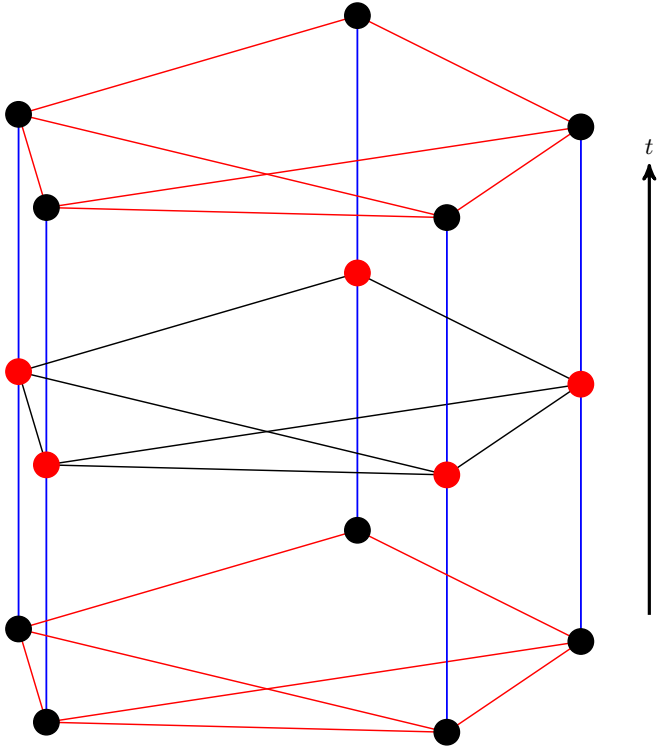
\begin{figure}
    \begin{center}
        \minipage{0.5\textwidth}
        \includegraphics[width=\linewidth]{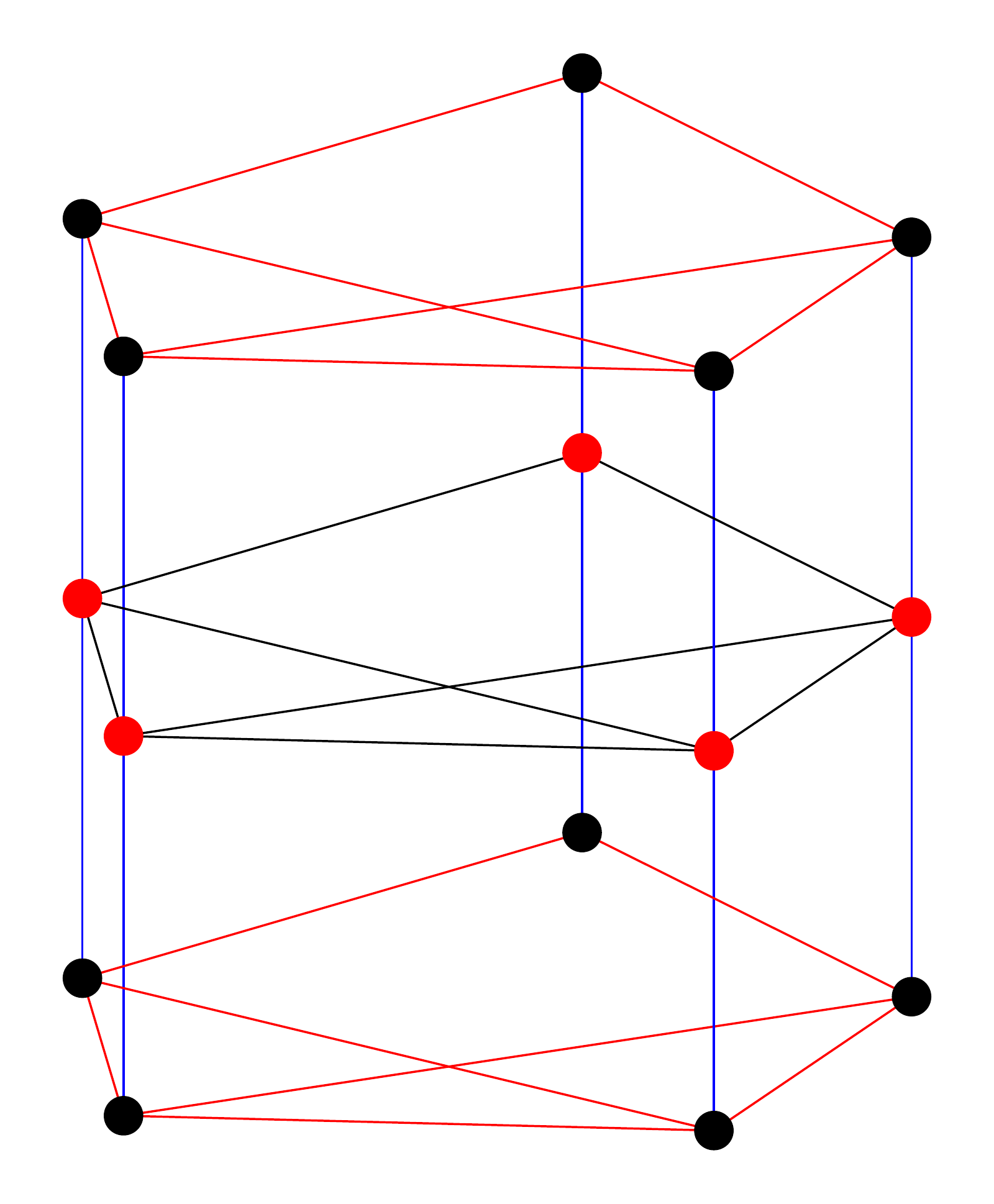}
        \endminipage
        \minipage{0.01\textwidth}
        \begin{tikzpicture}[scale=5, axis/.style={very thick, ->, >=stealth'}]
            \draw[axis] (0,-0.1) -- (0,1.1) node(yline)[above] {$t$};
        \end{tikzpicture}
        \endminipage
    \end{center}
    \caption{Five qubits evolve as time goes up on the page.  At $t=0$, commuting interaction terms (red edges) are applied between the qubits.  At $t=1$ local gates (red nodes) are applied to each of the qubits. }
    \label{fig:multiplex}
 \end{figure}

 \begin{remark}     An active edge (node) will specify if neighboring edges (nodes) can be active. As a general rule, non-commuting terms must be active on different layers.  
     \begin{enumerate}
        \item Qubits connect network layers by time propagation.
        \item Nodes of $S$ in each layer can be acted on by local gates.
        \item Edges of $S$ in each layer can be acted on by two-qubit gates. 
    \end{enumerate}
 \end{remark}
 As an example, consider the following multiplex network depicted in Fig. \ref{fig:multiplex}.    

    In the bottom layer a sequence of red edges correspond to the application of two-body gates.  In the next layer, the red (active) notes correspond to local rotation gates being applied to all the qubits. Finally, in the top layer the commuting two-body gates are applied.  The blue vertical edges represent time passing (going up on the page).  The hardware-efficient ansatz is exactly such a tiling.  
    
In terms of example circuits, we typically will be given a short repetitive sequence.  For example, here we consider qubits interacting on a ring.  A layer of local gates followed by a layer of two-body gates is as follows.     
    \begin{equation*}
        \Qcircuit @C=.75em @R=.75em {
            & \gate{U} & \qw & \qw & \ctrl{1}   & \qw        & \qw & \qw    & \qw & \qw        & \gate{R_Y} & \qw \\
            & \gate{U} & \qw & \qw & \gate{R_Y} & \ctrl{1}   & \qw & \qw    & \qw & \qw        & \qw        & \qw \\
            & \gate{U} & \qw & \qw & \qw        & \gate{R_Y} & \qw & \qw    & \qw & \ctrl{2}   & \qw        & \qw \\
            &          &     &     &            &            &     & \ddots &     &            &            &     \\
            & \gate{U} & \qw & \qw & \qw        & \qw        & \qw & \qw    & \qw & \gate{R_Y} & \ctrl{-4}  & \qw \\
        }
    \end{equation*}
    
    \bigskip
    
Where the local gates, $U$ are arbitrary and the two-body gates are controlled $Y$ rotations.  What is the maximal bipartite entanglement that such a circuit can generate when acting on a product state? 

To understand this question, let us consider a pure $n$-qubit state $\ket{\psi}$. 

\begin{definition} 
Bipartite rank is the Schmidt number (the number of non-zero singular values) across any reduced bipartite density state from $\ket{\psi}$ (i.e.~$\lfloor n/2 \rfloor$ qubits).  
\end{definition} 

\begin{remark} 
Rank provides an upper-bound on the bipartite entanglement that a quantum state can support---a rank-$k$ state has at most $\log_2(k)$ ebits of entanglement. 
\end{remark} 

\begin{definition}
An ebit is a unit of entanglement contained in a maximally entangled two-qubit (Bell) state. 
\end{definition} 

\begin{remark} 
A quantum state with $q$ ebits of entanglement (quantified by any entanglement measure) contains the same amount of entanglement (in that measure) as $q$ Bell states. 
\end{remark} 

\begin{remark} 
If a task requires $r$ ebits, it can be done with $r$ or more Bell states, but not with fewer.  Maximally entangled states in $\mathbb{C}^d\otimes \mathbb{C}^d$ have $\log_2(d)$ ebits of entanglement. 
\end{remark} 

Now we arrive at what we call a {\it quantum circuit combinatorial area law}.  It is the minimal depth circuit that possibly could saturate the bipartite entanglement with respect to any bipartition. 

\begin{lemma}
 Let $c$ be the depth of 2-qubit controlled gates in the $n$-qubit hardware-efficient ansatz.  Then the maximum possible number of ebits across any bipartition is 
    $$ \min \{ \left \lfloor{n/2}\right \rfloor, c \}$$
 \end{lemma} 
 
    
        
    
 \begin{example}[Combinatorial quantum circuit area law] 
    Minimal possible $c$ saturating specific graph (see Tab. \ref{tab:circ_depth}). 
 \end{example}

\section*{Popular Summary}   

What is the full application scope of contemporary (noisy) quantum information processing devices?  Although many applications have now been discovered, ranging from quantum approximate optimization, to the simulation of quantum systems and more recently quantum enhanced deep learning, many celebrated text-book quantum algorithms (such as Shor's quantum factoring algorithm) do not appear to admit a variant which functions on existing noisy quantum information processors. 

What then is the strict boundary between quantum enhanced information processing (without error correction) and full error corrected quantum computation? Moreover, what is the application scope and ultimate algorithmic capacity of noisy quantum information processing devices in the absence of error correction? 

In the present study, we show that the contemporary variational approach to quantum enhanced algorithms admits a universal model of quantum computation.  Hence we establish that universal quantum computation can be cast in terms of resources which are accessible using contemporary devices, which implies that the unknown limits of existing quantum enhanced noisy processors can now be better probed. The model maps universal quantum computation into that of state-preparation and the minimization of a (polynomial in the problem size) sequence of expected values. As variational methods are inherently agnostic to how  quantum states are prepared, this enables experimentalists to vary the accessible control parameters to minimize an external and iteratively calculated objective function to preform universal quantum computation. Although the absolute limitations of this model in the absence of error correction remain unknown, a robust method of error suppression could make the approach amenable to a wider scope of traditional text-book quantum algorithms.


\begin{table}[h!]
  \caption{Table. Minimal possible circuit depth $c$ possibly saturating bipartite entanglement with respect to different interaction geometries.}
  \label{tab:circ_depth}
  \centering
  \begin{tabular}{p{1.5cm}|c|c|c}
     & line & ring & gird \\
    \hline    \hline
    interaction geometry
    &
    \begin{minipage}{.25\textwidth}
      \includegraphics[width=1.\linewidth]{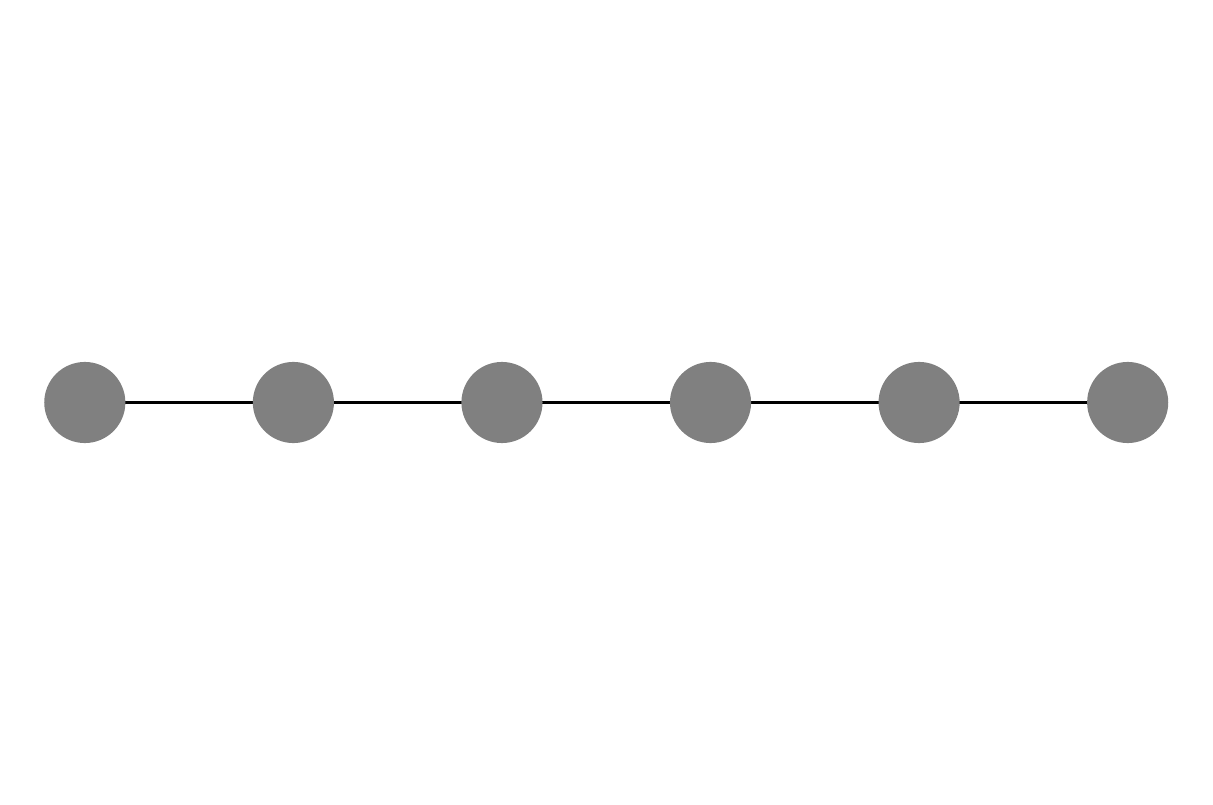}
    \end{minipage}
    &
    \begin{minipage}{.25\textwidth}
      \includegraphics[width=1.\linewidth]{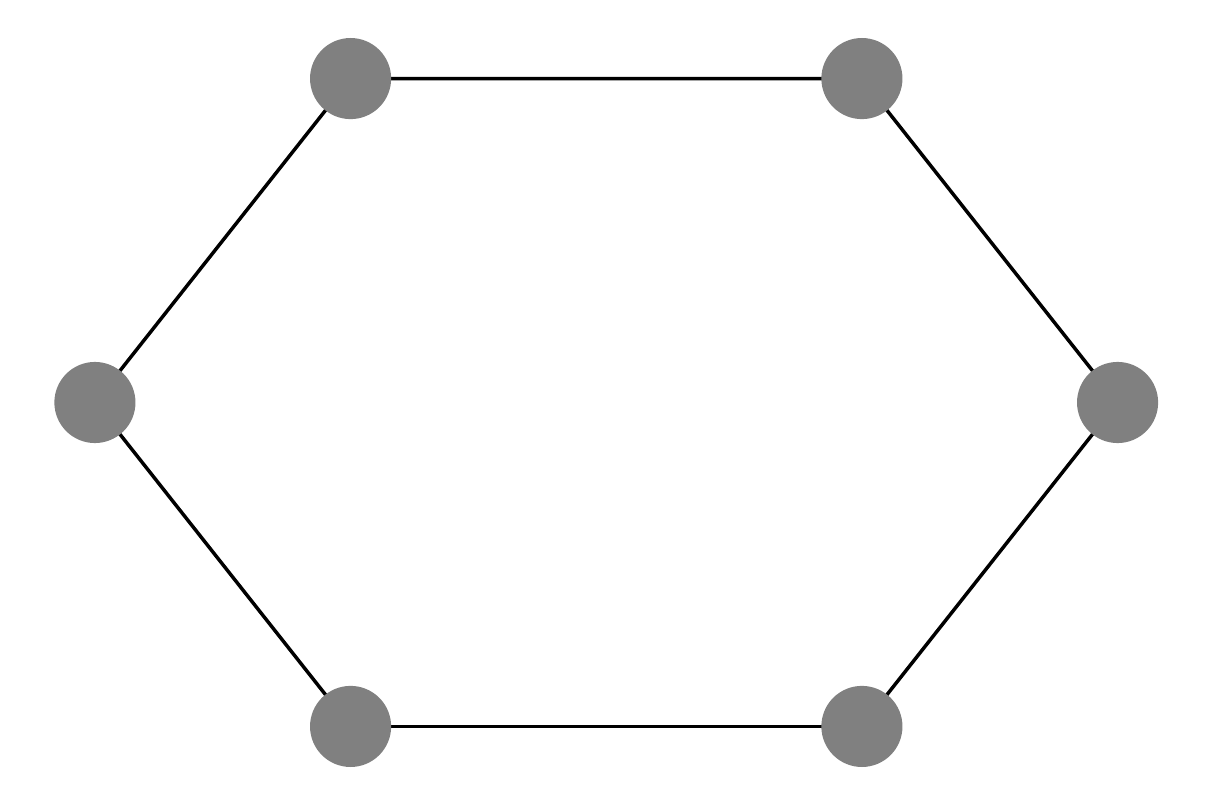}
    \end{minipage}
    &
    \begin{minipage}{.25\textwidth}
      \includegraphics[width=1.\linewidth]{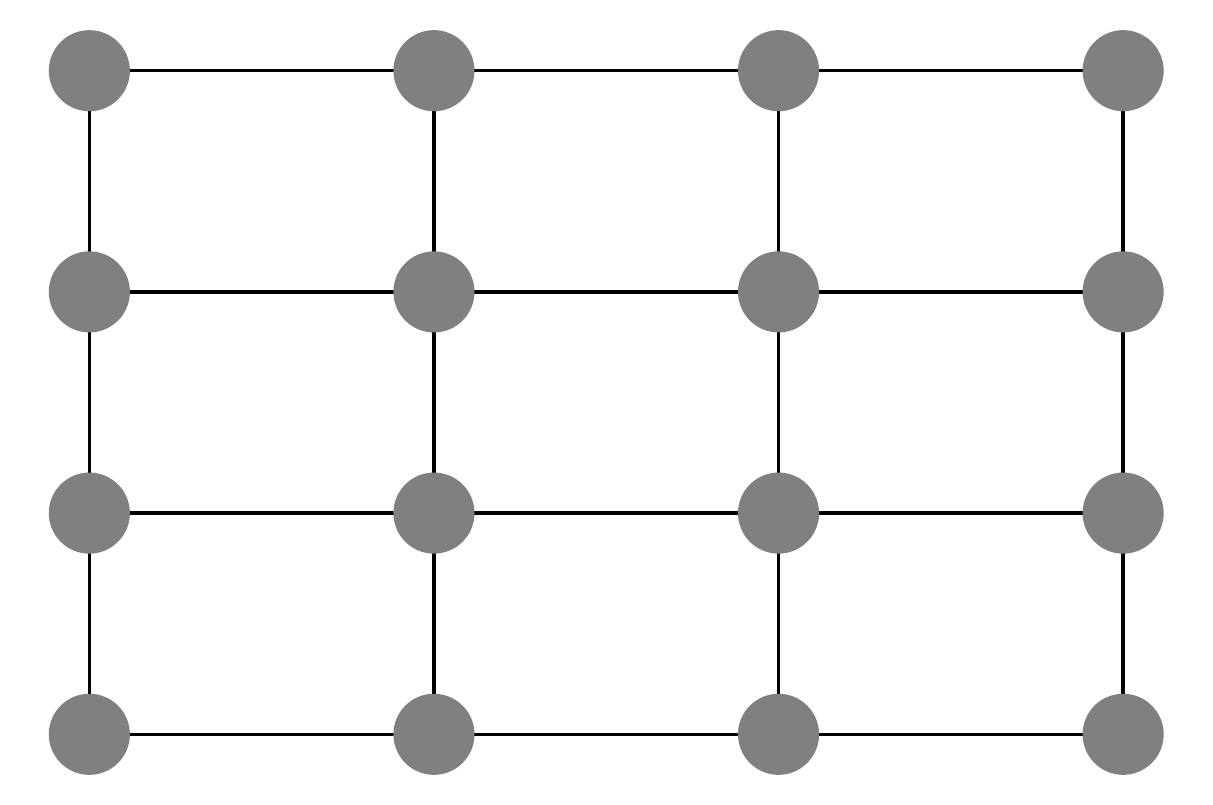}
    \end{minipage}
    \\
    \hline
    saturating depth & $c \sim n/2$ & $c \sim n/4$ & $c \sim \sqrt{n}/2$

  \end{tabular}
\end{table}

\end{document}